\documentclass[usegraphicx,useAMS,usenatbib]{mn2e}

\usepackage{times}

\title[Submillimetre Imaging Polarimeter]
{A Submillimetre Imaging Polarimeter at the James Clerk Maxwell
Telescope}
\author[J.S. Greaves et al.]
{ 
 J.S.\ Greaves$^{1,2}$, 
 W.S.\ Holland$^{1,2}$, 
 T.\ Jenness$^{1}$, 
 A.\ Chrysostomou$^{3,1}$, 
 D.S.\ Berry$^{4,5}$, 
\newauthor
 A.G.\ Murray$^{6,8}$, 
 M.\ Tamura$^{9}$, 
 E.I.\ Robson$^{1,5}$ 
 P.A.R.\ Ade$^{6,7}$, 
 R.\ Nartallo$^{6,7}$, 
\newauthor
 J.A.\ Stevens$^{10,2}$,
 M.\ Momose$^{11}$, 
 J.-I.\ Morino$^{12}$,
 G.\ Moriarty-Schieven$^{13,1}$, 
\newauthor
 F. Gannaway$^{6,7}$, 
 C.V. Haynes$^{6,7}$ \\
$^{1}$Joint Astronomy Centre, 660 N. A`oh\={o}k\={u} Place, University
Park, Hilo, Hawaii 96720, USA \\
$^{2}$UK Astronomy Technology Centre, Royal Observatory, Blackford Hill,
Edinburgh EH9 3HJ, UK \\
$^{3}$Division of Physical Sciences, University of Hertfordshire, College
Lane, Hatfield, Herts. AL10 9AB, UK \\
$^{4}$Department of Physics and Astronomy, University of Manchester,
Manchester, M13 9PL, U.K \\
$^{5}$Centre for Astrophysics, University of Central Lancashire, Preston,
   Lancs. PR1 2HE, UK \\
$^{6}$Department of Physics, Queen Mary, University of London, Mile End
Rd., London E1 4NS, UK \\
$^{7}$Department of Physics and Astronomy, Cardiff University, P.O. Box 913, 
Cardiff CF2 3YB, Wales \\
$^{8}$Moray College, Moray St., Elgin IV30 1JJ, UK \\
$^{9}$National Astronomical Observatory of Japan, Osawa, Mitaka, 
Tokyo 181-8588, Japan \\
$^{10}$ Mullard Space Science Laboratory, University College London, Holmbury
St. Mary, Dorking, Surrey, RH5 6NT \\
$^{11}$Institute of Astrophysics and Planetary Sciences, 
Ibaraki University, Bunkyo 2-1-1, Mito, Ibaraki 310-8512, Japan \\
$^{12}$Subaru Telescope, 650 N. A`oh\={o}k\={u} Place, Hilo, HI 96720, 
USA \\
$^{13}$National Research Council of Canada, Herzberg Institute of
Astrophysics, 5071 West Saanich Road, Victoria, BC, V9E 2E7, \\
Canada \\
}

\newcommand{\nat}{Nature}
\newcommand{\aap}{A\&A}
\newcommand{\mnras}{MNRAS}
\newcommand{\apj}{ApJ}
\newcommand{\apjs}{ApJS}
\newcommand{\pasp}{PASP}
\newcommand{\procspie}{Proc.\ SPIE}
\newcommand{\mum}{$\umu$m}

\begin{document}
\label{firstpage}
\maketitle

\begin{abstract}
A polarimeter has been built for use with the Submillimetre Common-User
Bolometer Array (SCUBA), on the James Clerk Maxwell Telescope (JCMT) in
Hawaii. SCUBA is the first of a new generation of highly sensitive
submillimetre cameras, and the UK/Japan Polarimeter adds a polarimetric
imaging/photometry capability in the wavelength range 350 to 2000~\mum.
Early science results range from measuring the synchrotron polarization of
the black hole candidate Sgr~A* to mapping magnetic fields inferred from
polarized dust emission in Galactic star-forming clouds. We
describe the instrument design, performance, observing techniques and data
reduction processes, along with an assessment of the current and future
scientific capability. 
\end{abstract}


\section{Introduction}

Linear polarization of submillimetre continuum radiation is associated
with optically-thin synchrotron emission, and with partial magnetic
alignment of elongated dust grains. In both cases, magnetic morphology can
be deduced from the directions of polarization vectors, and these
techniques are much less subject to confusion than in the optical and
radio regimes, where scattering and Faraday rotation dominate the
polarized signal. In the submillimetre, the limitations are that the data
are only sensitive to the net plane-of-the-sky magnetic field direction
within the telescope beam, and that there is no direct information on
magnetic field strengths. However, with the capability to detect levels of
linear polarization of 1\% or less, considerable magnetic {\it structure}
information can be obtained. The importance of magnetic fields is now
being realised in sources ranging from disks around young stellar objects
to the inner jets in active galactic nuclei
\citep[]{1999ApJ...525..832T,1998MNRAS.297..667N}. 

\begin{figure*}
\includegraphics[width=84mm]{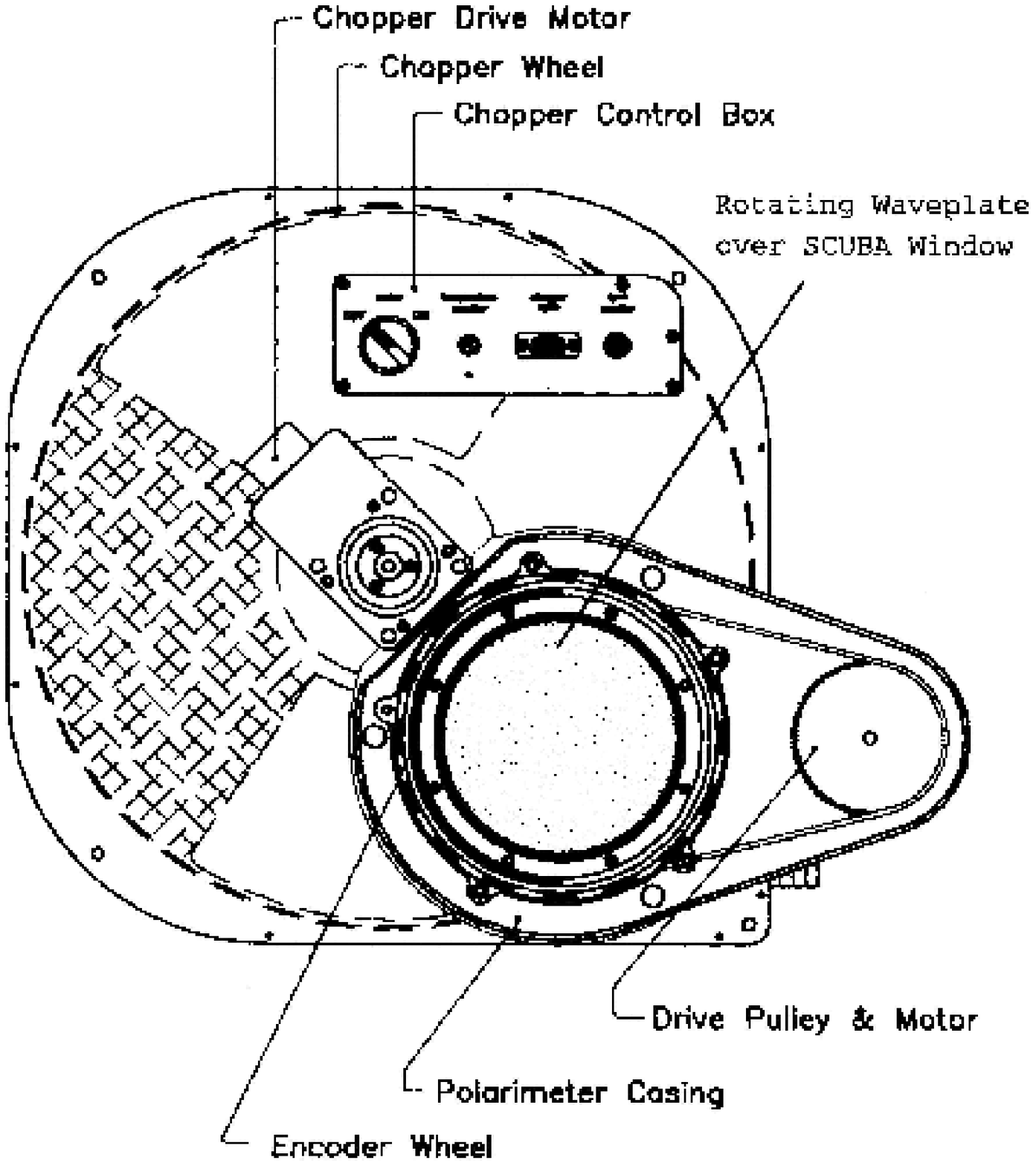}
\includegraphics[width=84mm]{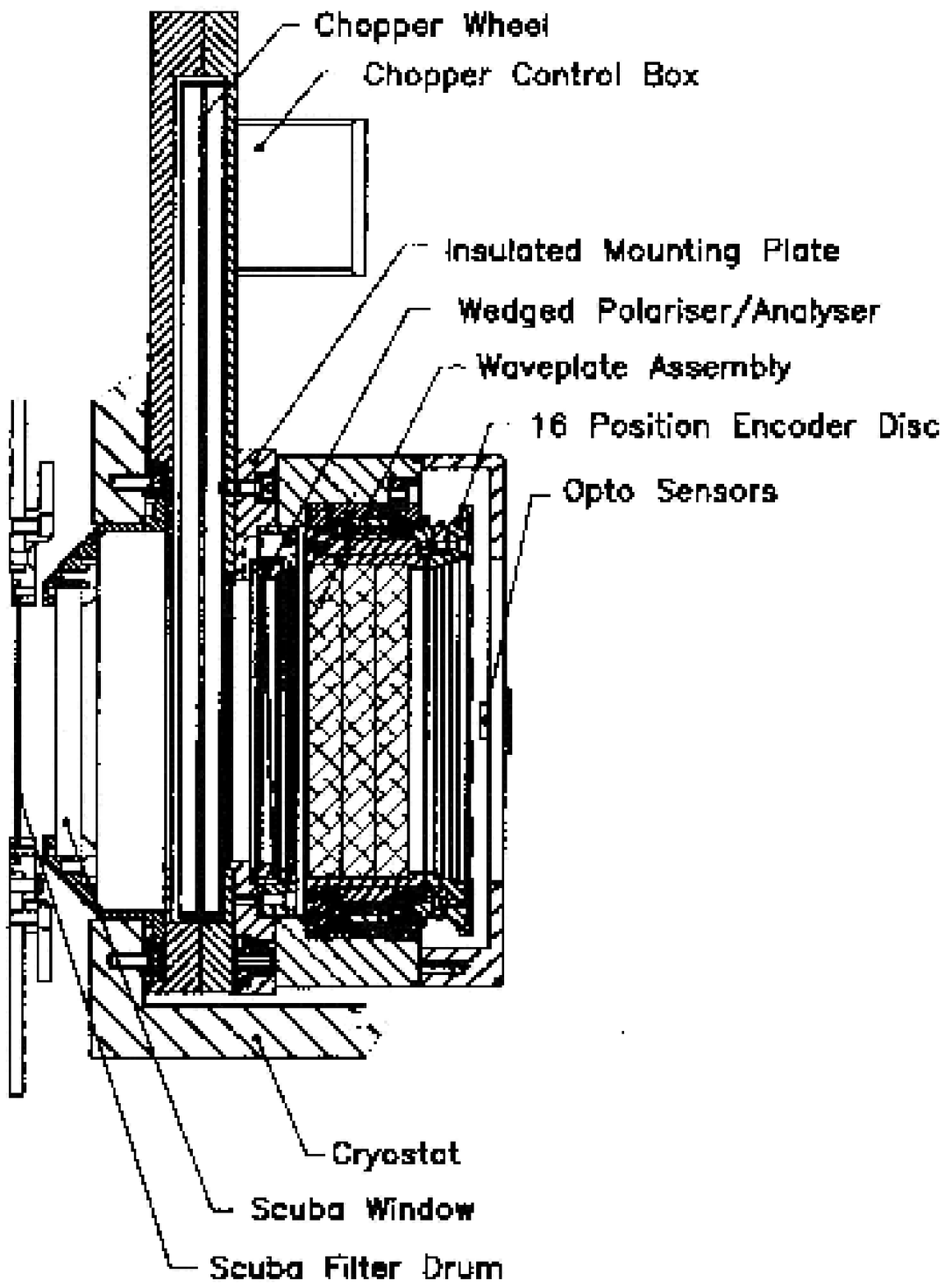}
\caption{Technical drawing of polarimeter (from \citet{1997fisu.conf..405M}). 
The physical dimensions of the polarimeter module are 30 by 10 cm (maximum 
length and thickness respectively). The waveplates are 12~cm in diameter; the 
thickness of individual half-wave quartz plates depends on wavelength but a 
representative value is 0.9~cm for a single plate optimised for 850~\mum\ 
observations and made of quartz with a submillimetre refractive index of 2.1.} 
\label{fig:diag}
\end{figure*} 

Observing techniques at submillimetre wavelengths have matured considerably
in the last decade, particularly with the introduction of the first
generation of continuum cameras. The chief difficulty for ground-based
observations is the high opacity and time variability of the Earth's
atmosphere, and these problems are critical for polarimetry. Atmospheric
absorption can be largely overcome by building telescopes at suitable high,
dry sites, such as Mauna Kea in Hawaii (summit altitude 4200m), and by
observing in frequency `windows' away from oxygen and water absorption
lines. Under the best conditions, the zenith transmission is then about 80\%
at a wavelength of 850~\mum, and as much as 40\% at shorter wavelengths
such as 450~\mum. The other problem affecting the quality of flux
measurements is atmospheric instability, especially emissivity changes on
short timescales of a few seconds.  This is a particular challenge for
polarimetric observations, which require measurements accurate to better
than 1\% over periods of minutes to hours, and necessitates sophisticated
techniques for removing sky-level changes.

Submillimetre polarimetry using single-pixel photometers began about a
decade ago \citep[e.g.,][]{1991MNRAS.249P...4F}, and a number of important
detections were made both in the (sub)millimetre
\citep[e.g.,][]{1993ApJ...411..708K,1997ApJ...480..255G,1999ApJ...511..812G}
and with complementary studies of warmer sources in the far-infrared
\citep[e.g.,][]{1989ApJ...345..802N}. These observations were extremely
laborious, but a major breakthrough has been made with the arrival of
sensitive submillimetre and far-infrared cameras. With the inclusion of
polarimeters --- typically based on rotating half-wave plate designs and
either internal or external to the cryogenically cooled instrument ---
polarization imaging of a wide variety of sources has now become feasible.
A complete introduction has been presented by \citet{2000PASP..112.1215H}.
In this paper, we describe observations and techniques with the new
imaging polarimeter\footnote{The SCUBA Polarimeter has been funded
  jointly by the UK Particle Physics and Astronomy Research Council and the
  NAOJ and JSPS of Japan. SCUBA was funded by the JCMT Development Fund supported
  by PPARC, the National Research Council of Canada and the Netherlands
  Organisation for Pure Research, and built by the Royal Observatory
  Edinburgh, now the UK Astronomy Technology Centre.} used with SCUBA, the
Submillimetre Common-User Bolometer Array \citep{1999MNRAS.303..659H} at the
James Clerk Maxwell Telescope (JCMT) on Mauna Kea.

\section{Instrument Design and Performance}

\begin{figure}
\includegraphics[width=84mm]{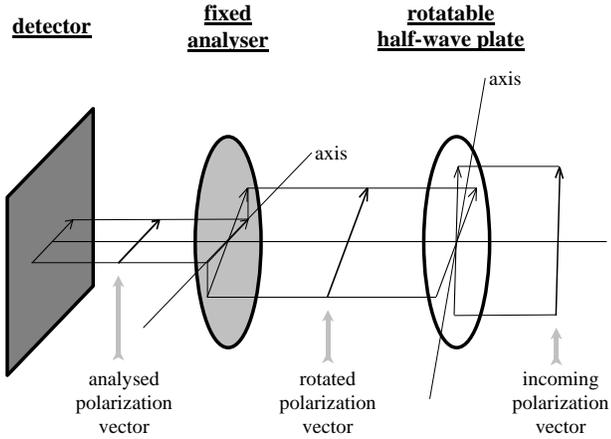}
\caption{Illustration of the effects of a rotating half-waveplate
and analyser \citep{sun223}.}
\label{fig:wplate}
\end{figure}

The general design of the polarimeter has been described by
\citet{1997fisu.conf..405M}. It is an external module that fits over the SCUBA
cryostat window, and comprises a rotating half-waveplate and a fixed
photo-lithographic grid (`analyser') (Figure~\ref{fig:diag}).  The waveplate
is stepped to a series of fixed angles, which has the effect of rotating the
incoming source plane of polarization, by twice the physical angle of
rotation. The grid, with a 10 micron line spacing (of etched copper on a 
mylar substrate), efficiently reflects one
plane of polarization that is fixed in the SCUBA reference frame.  The net
effect (Figure~\ref{fig:wplate}) is to select out a fixed component of a
rotated linearly polarized signal, so that the SCUBA bolometers see a
modulated signal of the form \begin{equation} S(\delta) = 1/2 [I_u + I_p (1 +
cos (4\delta - 2\theta))].  \label{eqn:sig} \end{equation} Here $I_u$ is the
unpolarized intensity, $I_p$ describes the modulating polarized component 
above the minumum observed signal $I_u$, $\delta$ is the waveplate angle, and 
$\theta$ is the position angle of the polarization in a suitable co-ordinate 
frame. The factor of $1/2$ arises because one plane of polarization is 
reflected by the analyser;  this contrasts with cameras designed specifically 
for polarimetry, which typically use a waveplate inside the instrument and 
measure both planes of polarization using two separate detector arrays
\citep{1998ApJ...504..588D}.

\subsection{Waveplate design}

The waveplates are made of birefringent quartz, cut to a thickness to
retard one plane of polarization by half a wavelength relative to the
orthogonal plane, on emerging from the plate. Since SCUBA has filters for
observing at a wide range of wavelengths, from 0.35 to 2 mm, an {\it
achromatic} design was adopted following the Pancharatnam method 
\citep{1997fisu.conf..405M,1981SPIE.307..120T}. This uses a sandwich of an odd 
number of individual half-waveplates whose `fast' axes of polarization are 
offset by $\pm 60^{\circ}$. The effect is a combined plate whose fast-axis 
direction ($\phi$) is slightly wavelength-dependent, but with very good 
polarization modulation efficiency (PME) over a broad range of wavelengths. 
The plates were wax-bonded and then held in place in mounting rings. A good 
transmission ($\eta$) was achieved using single-layer polypropylene 
coatings to reduce reflections at the plate surfaces. 

\begin{table}
\centering
\caption{Waveplate properties (symbols defined in the text). The input
signal was produced with a second analyser to make a pure linear
polarization from an astronomical source (generally Saturn). The PME is
a function of waveplate angle with respect to the incoming plane, but
minimum and mean values differ by no more than 8\%. The $\phi$ values in
brackets refer to observations after late 1998, when the
photometric plate was re-mounted; the 1350~\mum\ value was measured
and the shift at 2000~\mum\ was inferred. } 
\begin{tabular}{cccc} 
$\lambda$  & mean PME & $\phi$    & $\eta$ \\
(\mum) & (\%)     & (degrees) & (\%) \\
\hline
\multicolumn{4}{l}{{\it `Array' waveplate}} \\
350 & 96 & +9 & --- \\
450 & 97 & +7 & $\sim$60\% \\
750 & 99 & +1 & --- \\
850 & 97 & --5 & $>$85\%\\
\multicolumn{4}{l}{{\it `Photometric' waveplate}} \\
 450 &  95 & +7 & --- \\
 850 &  94 & +6 & --- \\
1350 &  99 & +18 (+8) & $>$85\% \\
2000 &  93 & +3 (--7) & $>$85\% \\
\hline
\end{tabular}
\end{table}

For the SCUBA polarimeter, two achromatic designs were used: one with
three quartz plates to cover the 1.1, 1.35 and 2 mm bands and one
with five plates for the 350, 450, 750 and 850~\mum\ bands. Because
SCUBA has single photometers at the millimetre wavelengths and full
imaging arrays at the submillimetre wavelengths, the plates are known
as `photometric' and `array' for convenience. 

\subsection{Performance on the JCMT}

The performance of the waveplates on the telescope was first measured in
October 1997, and the results are given in Table~1. Excellent PME values
were measured, close to 100\% at all wavelengths, and it was even found
possible to operate the photometry plate in second order (3$\lambda$/2) at
the array wavelengths. (This has advantages for complex observing
programmes, because the polarimeter need not be removed to change the
plate.) Nominally, measured polarization percentages should be divided by
the PME to obtain the true values, but in practice this small correction
is usually neglected. The transmission properties met the specifications
for a single-layer coating, except at the shortest observing wavelengths,
and it is suspected that re-coating and bonding of the plate might be
necessary to improve this.

\begin{table*}
\centering
\caption{Percentage and position angle ($p,\theta$) of the main-beam
instrumental polarization (not corrected for PME or $\phi$); for the array
data the uncertainties are the standard error of the set of individual
measurements. Observations were made using several planets; note that Saturn
may be slightly polarized due to scattering by the rings (with an best-fit
value of 0.5\% at 850 \mum); however this effect is greatly reduced in data
averaged over many hour angles. Saturn is also larger than the beam size at
all the submillimetre wavelengths, so these data include some degree of
sidelobe polarization. } 
\begin{tabular}{ccccc}
$\lambda$  & p(IP) & $\theta_0$(IP) & source & epoch\\
(\mum) & (\%)  & (degrees)      & & \\
\hline
\multicolumn{5}{l}{{\it Array waveplate (central array bolometer)}} \\
350 & 0.82 $\pm$ 0.13 & 100 $\pm$ 4 & Saturn & Oct 1997 + Aug 1999 \\
450 & 3.51 $\pm$ 0.34 & 113 $\pm$ 3 & Mars/Uranus & 1997--1999 (5 dates) \\
750 & 1.16 $\pm$ 0.04 &  97 $\pm$ 1 & Saturn & Oct 1997 + Aug 1999\\
850 & 1.09 $\pm$ 0.06 & 161 $\pm$ 2 & Mars/Uranus & 1997--1999 (25 dates)\\
\multicolumn{5}{l}{{\it Array waveplate (off-centre bolometers)}} \\
450 & 3.26 $\pm$ 0.23 & 100 $\pm$ 2 & Saturn & Jul 1998 (6 bolometers only) \\
850 & 0.92 $\pm$ 0.05 & 163 $\pm$ 2 & Mars/Uranus/Saturn & 1997--1999 (4 dates, all 36 bolometers) \\
\multicolumn{5}{l}{{\it Photometric waveplate}} \\
1350 & 1.72 $\pm$ 0.11 & 166 $\pm$ 2 & Uranus & 1997--1999 (7 dates)\\
2000 & 1.34 $\pm$ 0.09 & 170 $\pm$ 1 & Saturn/Uranus & Oct 1997 \& Aug
1999\\
\hline
\end{tabular}
\end{table*}

\subsection{Reproducibility}

An important part of the commissioning test was to confirm that the
results agreed with previous polarimetric observations, from the JCMT
and elsewhere. Tests included both low and high polarization sources: 
for example, 
\begin{itemize}
\item{850~\mum\ polarization of the DR21 cloud core was
measured to be 1.70 $\pm$ 0.27\% at $22 \pm 4^{\circ}$, whereas Minchin \&
Murray (1994) found 1.8 $\pm$ 0.3\% at 17 $\pm$ 4$^{\circ}$ at 800~\mum}
\item{a flux peak in the Crab Nebula was detected at 1350~\mum\ with 25.9
    $\pm$ 0.6\% polarization at 148 $\pm$ 1$^{\circ}$, while
    \citet{1991MNRAS.249P...4F} obtained 25 $\pm$ 2\% at 144 $\pm$ 2$^{\circ}$
    at 1100~\mum.}
\end{itemize} 
These results were from previous photometric polarimetry at the JCMT, and are
for dust and synchrotron emission respectively. Since the previous instrument
had no imaging capability, reproducibility in this mode relied on data from
other telescopes: for example, our 850~\mum\ polarization map of OMC1
\citep{2000A&A...356.1031C} is very similar to the 350~\mum\ results of
\citet[][his Figure~5]{1998ApJ...493..811S}. The position angles of dust
polarization should be wavelength-independent provided that the telescope
beams sample the same grain population and line-of-sight
\citep[]{2000PASP..112.1215H}. 

\subsection{Sensitivity limits}

The sensitivity for polarization measurements is determined essentially by the
sensitivity of SCUBA, with additional effects of the transmission through the
polarimeter and the limiting accuracy of corrections for sky fluctuations. The
critical quantity is the polarized flux: the product of percentage
polarization and mean flux. With the previous JCMT polarimeter used on a
single-bolometer detector, UKT14, the smallest polarized flux that could be
measured was about 100 mJy at 800~\mum\ \citep{1997fisu.conf..405M}.  This is
equivalent, for example, to p~=~1\% in a 10 Jy source, such as one of the
brightest Galactic protostars.

With the SCUBA polarimeter, the expected sensitivity can be derived by
considering the magnitude of the fractional polarization, as obtained by
subtracting signals at two waveplate angles and dividing by their
sum\footnote{Two waveplate angles 45$^{\circ}$ apart will have the largest
and smallest observed signals, $I_{max}$ and $I_{min}$, and it can be
demonstrated from Equation~\ref{eqn:sig} that $(I_{max} - I_{min}) /
(I_{max} + I_{min})$ is equivalent to the polarization fraction, $p =
I_p/(I_u + I_p)$. For convenience we re-write $p = (R - 1) / (R + 1)$
where $R = I_{max}/I_{min}$, and differentiate this to give $dp = 2 / (R +
1)^2 dR$. For small polarizations $R \approx 1$ so $dp \approx dR/2$. Then
as the two $I$ measurements are independent, $dR/R = \sqrt{2} dI/I$ and
combining the last two expressions, $dp = dI/I \times 1/\sqrt{2}$ in
the limit $R \rightarrow 1$.}. The error on the polarization fraction is
$1/\sqrt{2}$ times the fractional intensity error, so half as much time is
needed as would be for a measurement of $I$ alone. However, this is
cancelled by a factor of twice as much time that arises because half the
photons are lost to reflection off the analyser. Finally there is another
time factor of four because four waveplate positions are needed to obtain
the two orthogonal linear polarization components (Stokes parameters).
This determines the signal-to-noise on the polarization percentage,
$\sigma_p$, for a given integration time $t$: \begin{equation} t = 4 /
\eta \times N \times ({\rm NEFD} \sigma_p / p F)^2 \label{eqn:tint}
\end{equation} where NEFD and $F$ are the noise equivalent flux density
and source flux in mJy/$\sqrt{\rm Hz}$ and mJy respectively, and $p$ is
the fractional polarization (e.g. 0.01 for 1\%). The factor N is the
inverse of the fraction of time spent on one spatial point, so it is 1 for
photometric polarimetry, 4 for imaging polarimetry at beam width spacings,
and 16 for a Nyquist-sampled polarization map.\footnote{The SCUBA beams
are spaced two beam widths apart on the sky, so that at beam width
resolution a bolometer is effectively looking at a particular point for
1/4 of the time. Since the array is undersampled, the telescope secondary
mirror is `jiggled' to fill in the image; the polarization maps are
usually sampled every 6$''$ but smoothed to about beam width spacing to
improve $\sigma_p$. For an extended source, smoothing with e.g. a 
Gaussian function includes more flux and hence reduces $t$ to obtain a 
given $\sigma_p$.}

At the primary observing wavelength of 850~\mum, the NEFD is 70 mJy/$\sqrt{\rm
Hz}$ under the best conditions, and a 1 hour integration would give a
3$\sigma$ detection for a polarized flux $p \times F = 15$~mJy in a
beam-spaced map, or 7.5 mJy for photometry. This signal-to-noise is sufficient
to determine magnetic field directions to 0.5 radians / 3
\citep{1993A&A...274..968N}, or about $\pm 10^{\circ}$, and the sensitivity is
more than an order of magnitude better than with the UKT14 polarimeter. The
faintest sources actually detected have had fluxes of about 0.2 Jy per beam
and polarizations of a few percent;  scientifically, large samples of
protostars, faint Galactic clouds and AGN can now be detected. Also, types of
sources that were previously impossible can now be studied, including
asteroids, pre-stellar cores and nearby starburst galaxies.

\section{Instrumental limitations}

\subsection{Main-beam instrumental polarization}

The absolute accuracy of any polarization measurement depends on how
reliably instrumental effects can be subtracted. For the SCUBA polarimeter
the main problem is instrumental polarization (IP), which is dominated by
the woven Goretex windblind through which the JCMT observes. The thread
spacing, which is approximately 0.5~mm, is slightly different in the
vertical and horizontal directions and this affects the relative
transmissions, particularly at a wavelength of 450~\mum\ which is closest
in size. The solution would be to either roll back the windblind (a
half-hour labour-intensive operation that is only possible on rare
occasions of very low windspeed) or add a second orthogonal sheet of
Goretex in the beam to induce IP cancellation. The latter has been tried
and reduced the IP by a factor of at least three at 450~\mum; further
experiments are ongoing but this is not yet the standard observing mode.

The IP in the main beam of the telescope has been measured using planets,
assumed to be unpolarized \citep[e.g.,][]{1990PASP..102.1064C}, and the
results are listed in Table~2. At the primary wavelength of 850~\mum\ the
measured IP values are very stable and can be
subtracted to an accuracy of about $\pm 0.25$\%, the measurement
uncertainty for individual bolometers. Global differences, such as
off-axis effects in the three rings of bolometers that form the hexagonal
close-packed array, are 0.1\% or less, and for the central array bolometer
(used for photometric polarimetry) the measurement error has been reduced
to about $\pm 0.06$\%. Fewer data are available at other wavelengths but
generally a very accurate measurement can be made during a particular
observing run. Values tend to be stable over time, taking into account
that the angles must be corrected for telescope elevation. The observed
$\theta$ is $\theta_0$ plus elevation, due to the alt-az telescope
mounting and the fact that SCUBA is aligned with the elevation axis but is
fixed on the Nasmyth platform. This angle relation is quite reliable, with
deviations of around $\pm 15^{\circ}$ at extreme elevations, for the 850~\mum\
pixels at the array edges that are furthest from the optical axis. 

\subsection{Sidelobe polarization}

\begin{figure}
\includegraphics[width=84mm,clip=]{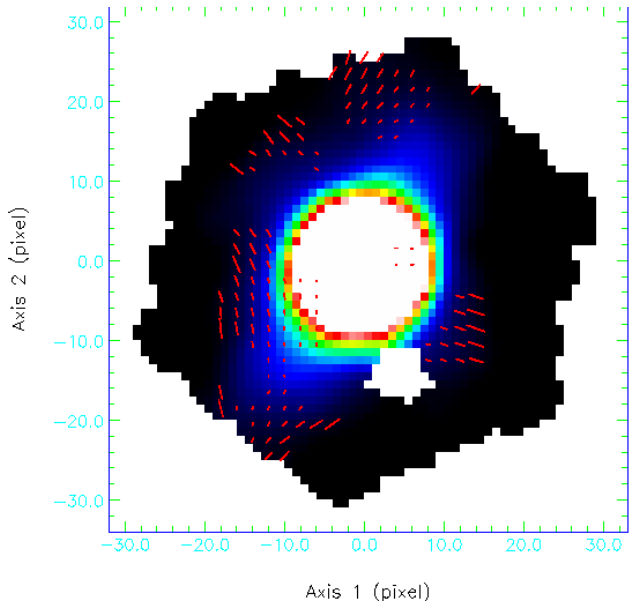}
\caption{Typical 850~\mum\ map of polarization measured around Jupiter, due
to sidelobe IP. Vectors shown range from 0.8\% to 15\% (with criteria 
$p/\delta p > 2$ and $\delta p < 3$\%) and are plotted 
every $6''$ (2 pixels). Main-beam instrumental polarization has been
subtracted. Data are from January 2000, using a 180 arcsec chop of the
secondary mirror in the azimuth direction.} 
\label{fig:jup}
\end{figure} 

The sidelobes of the telescope beam have low intensity-amplitudes (e.g. 
$<1$\% at 850~\mum\ at $> 40''$ from Jupiter in Figure~\ref{fig:jup}), but
different reflection and scattering effects for the two incoming planes of
polarization can still produce instrumental `sidelobe polarization'.  This
needs to be considered when observing extended sources, or faint objects
next to much brighter ones, as the sidelobe IP can be considerable, and is
additional to the main-lobe IP that has been measured using a compact
planet.

In theory, a polarization map of an extended unpolarized source (generally
Jupiter), made with the same primary dish surface and secondary mirror
chop throw as the source observations, can be used for an exact
subtraction of sidelobe IP. An example of such a map is shown in
Figure~\ref{fig:jup}, after subtraction of the standard main-lobe
instrumental polarization.  However, in practice it is often easier simply
to estimate the error induced by the sidelobe effects. The most extreme
case is seen with large chop throws of the secondary mirror (the maximum
reccomended is $\pm 180''$) in a non-azimuthal direction. In this case,
there are residual vectors towards about half of Jupiter's disk, with
magnitudes up to 3\%, although when {\it averaging} over this area
(diameter of 40 arcsec) the net effect is only 0.25\%. For smaller or
azimuthal chop throws, the net residual is similar or lower and only about
1 in 10 individual points has a significant vector.  Thus if the observer
is interested in average polarization of an extended source, the error is
quite small, but more careful calculations are required for data on a
single map point in an extended region.

Additional problems arise if trying to extract source polarization when
there is a much brighter object nearby in the field. In
Figure~\ref{fig:jup}, there are vectors averaging $\sim 6$\% around Jupiter,
although the largest tend to be at large radii where the power in the beam
is low. A good general method is to set a threshold of {\it source}
polarization percentage, above which sidelobe IP is not a serious
contaminant, and to base this threshold on the relative polarized flux
contributions. For example, if the instrumental polarized flux is half that
of the source polarized flux, then in the worst case the two effects have
orthogonal Stokes parameters, and the error in $\theta$ for the source will
be $1/2 \tan^{-1} 0.5$ or only 13$^{\circ}$. (If the effects are parallel,
then the error will instead be in $p$, which is generally of less scientific
importance.)

Taking into account the beam power profile, this criterion for the
minimum believable source polarization percentage can be written
\begin{equation}
p_{crit} \geq 2 \times p_{sl} P_{sl} (F_{sl}/F)
\label{eqn:pcrit}
\end{equation}
where the three terms on the right-hand side are the IP at the relevant
point in the sidelobe, the power here relative to the main beam, and
the ratio of the flux of the object in the sidelobe to the flux in the
main lobe. Thus sidelobe polarization can often be non-critical if the
first two terms are small, even if there is a bright object in the
sidelobe ($F_{sl} >> F$). As a quantitative example, \citet{2002A&A..PNE} 
observed polarization in the jets of the proto-planetary nebula CR~2688, at 
positions offset 15$''$ from the central star. The ratio $F_{sl}/F$ was 3.6, 
the average $p_{sl}$ was 3\% and the power at the edge of the main beam was 
$P_{sl} \approx 0.06$. Using Equation~\ref{eqn:pcrit}, the polarization of the 
southern jet was  found to be believable because the measured value of 1.9\% 
exceeded $p_{crit}$ evaluated at 1.3\%. At larger offsets, the critical 
product of power and IP tends to be smaller: for example the values $< 0.01$
and $\leq 15$\% (Figure~\ref{fig:jup}) give a smaller product than in the 
CRL~2688 case. The most critical quantity in individual sources is thus the 
flux ratio: very high contrasts will mean that faint regions need to be 
highly polarized to give usable results. 

\section{Observing Techniques and Data Reduction}

Data acquisition and reduction are based on extensions to standard
techniques used at the JCMT; more details of the standard procedures are
given in \citet{1998SPIE.3357..305H,1999MNRAS.303..659H},
\citet{1998adass...7..216J} and \citet{2000irsm.conf..205J}. For
polarimetry, standard imaging or photometry observations are made at a set
of waveplate positions (normally 0, 22.5, 45, 67.5, 90...
337.5$^{\circ}$), and these data are fitted to detect the sinusoidal
signal modulation from rotating the incoming polarization vector. Using a
complete cycle is intended to eliminate other modulations such as
reflections off the waveplate. An alternative to fitting is direct
subtraction of data taken at different angles, as for example, $p \;
cos(2\theta)$ can be found from $(S(45) - S(0)) / (S(45) + S(0))$ and $p
\; sin(2\theta)$ from $(S(67.5) - S(22.5)) / (S(67.5) + S(22.5))$
(Equation~\ref{eqn:sig}); this method is entirely valid and gives both
Stokes parameters directly, but is less often used. For an excellent
introduction to polarization observing procedures and limitations, see
also \citet{2000PASP..112.1215H}.

\subsection{Observing modes}

Three observing modes have been adapted for use with the polarimeter.  These
are {\it photometry}, where a single bolometer observes a fixed point; {\it
jiggle-mapping}, where the secondary mirror is jiggled to obtain an image of
one SCUBA field-of-view, using all the bolometers in an array; and {\it
scan-mapping}, where the telescope scans across a source and the data from all
bolometers are reconstructed to give large images. All of these modes use 
chopping of the secondary mirror to remove the sky emission; photometry and
jiggle-mapping also nod the source between left and right beams halfway
through an observation to minimise sky gradients. For polarimetry, the only
difference is that a complete standard observation is made at each waveplate 
angle before moving on to the next angle.

The first two polarimetry modes are fully commissioned, and the third has
been proven experimentally but is not yet fully integrated. All six of the
available SCUBA wavelengths have been used, but only a small number of
observations have been made with the 350/750~\mum\ filter combination or
using the 2~mm photometric pixel (less than 1\% each of the total number of
observations). Thus the best quantified performance is at 850 and
1350~\mum; 450~\mum\ data are obtained simultaneously with 850~\mum\ but
are often not usable due to poor atmospheric transmission.  Scan-mapping has
been tested only at 850~\mum. 

The photometry mode is most suited to sources with accurately known
positions that are smaller than the telescope beam width (8, 15 and 24
arcsec at 450, 850 and 1350~\mum). The off-source bolometers remain
`switched-on' and can be used to subtract residual sky signals (not
possible with the individual photometric pixels at 1350 and 2000~\mum).  
Typically the secondary mirror chops between on- and off-source with a 1
arcmin throw, to allow subtraction of the sky DC-level, and 8 seconds of
data are obtained at each of the 16 waveplate positions. The chop takes
place at a rate of 7.8 Hz, and the telescope nods so that the source moves
from the left beam to the right beam after 4 seconds. A small grid of
points can be used rather than a fixed point but this in not generally
done.  Integrations of up to about 20 seconds can provide higher
signal-to-noise in stable conditions, but generally many short
integrations are co-added.

The jiggle map mode is typically used for sources bigger than the beam but
smaller than the 2.3 arcmin array field-of-view in at least one dimension.  
If the source is larger, two effects can corrupt the polarization
measurement \citep{2001ApJ...562..400M}. Firstly, the maximum recommended
chop throw is 3 arcmin, so sources larger than this will be
`self-chopped'. In practice sky noise subtraction requires several
emission-free bolometers, so the setting of the zero level will be
inaccurate if the source dimensions approach 3 arcmin; an incorrect
zero-level changes $I_u$ and so alters the percentage polarization.
Secondly, if there is emission in the reference, i.e.  nominally
off-source, beams it may well be polarized and this can corrupt the
on-source polarized signal in both magnitude and direction. A limit can be
estimated where the reference polarization is not known, but this may not
set useful constraints: for example, faint outer regions in some dust
clouds can be up to $\sim 15$\% polarized
\citep[e.g.,][]{2001ApJ...562..400M}.

A 16-point jiggle sequence with the secondary mirror gives an 850 (or 750)  
\mum\ image with a point every 6 arcsec, so is slightly better than Nyquist
(half-beam) sampled. The simultaneous data at 450 (or 350)  \mum\ are
undersampled, since the points are only a little better than a beam width
apart, and a 64-point jiggle sequence with 3 arcsec sampling is required if
these data are of importance. The simpler jiggle takes 32 seconds to complete
at each waveplate position, and the more complete jiggle takes 128 seconds, so
the former is preferable for greater sky stability.

The scan map mode has been tested at 850~\mum\ only, on a few sources with
sizes of about 6--10 arcmin. The technique used is to scan the telescope
across the source in any suitable direction, while chopping in RA or Dec. 
Using two different chop throws in each direction, for a total set of
four, has been found to give good reconstructed images; further details of
this `Emerson II' technique are given by \citet{2000irsm.conf..205J}. It
is important to scan off-source so that a true baseline can be established
and set to zero in each individual waveplate image, before the
polarization reduction is done. 

The scan rate used is 48 arcsec per second, so sampling every 1/8th second
gives points 6 arcsec apart, the same as in jiggle mapping.  The minimum
number of waveplate positions needed to deduce the Stokes parameters is
four, so scan map polarimetry is rather time-consuming, for example,
taking 30 minutes for one dataset on a 400 arcsec square area. This
compares to three minutes or less for a set of Stokes parameters in the
other modes, so scan map polarization measurements require more stable
conditions.

\subsection{Data reduction techniques}

Two separate software packages are used for the data reduction: one for
photometric polarimetry \citep[SIT:][]{1995PhDT........54N} and one for
imaging polarimetry \citep[POLPACK:][]{sun223}. The reduction philosophy
is the same in both cases, and in fact POLPACK can handle single-pixel
`images', while SIT is more optimised for inspecting the results of
individual waveplate cycles on a single point. 

The raw data are reduced \citep{2002MNRAS..TJCAL} by extracting the
chopped signals, flatfielding, and correcting for atmospheric extinction
using a skydip. This is a series of measurements of sky emission at
different elevations, made either with the JCMT at the observing
wavelength, or taken from the 1.3mm database recorded at the adjacent
Caltech Submillimeter Observatory (CSO) and extrapolated in wavelength. 
The next refinement is to subtract rapid changes in sky emission (to a
greater accuracy than achieved simply by chopping: 
\citep{1999MNRAS.303..659H}) by shifting the mean level using blank
bolometers in the array, or in scan-mapping, subtracting smoothed data
seen by each bolometer over 2 seconds. This `sky noise' is spatially
correlated over the array diameter \citep*{1998SPIE.3357..548J}. 

The main lobe instrumental polarization is then removed on a 
bolometer-by-bolometer basis using the expression: 
\begin{equation} 
S_{corrected} \approx \frac{S_{measured}}{1 + p_{IP}(e) \cos(4\delta -
2\theta_{IP}(e))}
\label{eqn:remipapprox} 
\end{equation} 
where $p_{IP}(e), \theta_{IP}(e)$ are the percentage and direction of
instrumental polarisation at elevation $e$ (a constant percentage is
generally assumed). This is an approximation of
\begin{equation} 
S_{corrected} = S_{measured} - S_{mean} p_{IP}(e) \cos(4\delta -
2\theta_{IP}(e)) 
\label{eqn:remip} 
\end{equation} 
and is valid for instrumental polarizations of up to a few percent. The
former expression is prefered as the mean flux level is not easily 
determined prior to regridding --- the bolometer jiggles to different source 
positions and thus only part of the data stream refers to a particular 
spatial point. Also, not exactly the same point is seen by this bolometer for 
the next waveplate angle, because of sky rotation (SCUBA does not use an 
image de-rotator). 

In imaging mode the data are then regridded to a rectangular array, removing
pixels from the edge that are affected by regridding edge effects in order
to simplify mosaicing. For scan map data, a Fourier deconvolution is used to
remove the chop signature and combine the four different chopped maps
\citep{2000adass...9..559J}. 

The Stokes parameters I, Q, U are then extracted by fitting the signal in
each pixel as a function of waveplate angle; errors are supplied by
comparing signals at identical or equivalent waveplate positions (e.g. 0,
90, 180 and 270$^{\circ}$; see Eq.~\ref{eqn:sig}\footnote{Scan map
observations are sufficiently time-consuming that only the angles $\delta
= 0, 22.5, 45, 67.5^{\circ}$ may be used, in which case the same angle in
different cycles should be compared.})  In SIT, the Q and U are then
listed and statistical tests can be used to reject anomalous fits before
calculating averages and deducing $p,\theta$.  In POLPACK, an I, Q, U
data-cube is generated and a catalogue of vectors produced corresponding
to all the image pixels. When combining different data sets (a set
contains 16 waveplate positions) it was determined that the best results
were obtained by coadding the IQU cubes rather than adding more data to
the fit. As more data were added to the fit the signal-to-noise did not
increase in the expected way, suggesting that it was susceptible to
DC-level differences amongst the observations.

Various binning and selection procedures can be used to improve the
polarization signal-to-noise. Typical criteria are $p/\delta{p} > 3$ (i.e.
$\delta\theta < 10^{\circ}$) and an upper limit such as $\delta{p} < 1-2$\%
to eliminate biased data. Because the waveplate angles increase step by
step, a linear drift in the sky transmission can force, for example, a
positive Q and U, and these spurious fits tend to produce a high $p$,
especially in regions of low I. Thus a $\delta{p}$ criterion can eliminate
such data with a large scatter between good and inaccurate fits. For
example, four measurements with similar $\theta$ but $p = 3, 3, 3, 9$\%
would give an average $p$ of 4.5\% and a standard error of 1.3\% (hence
$p/\delta{p} = 3.5$). This dataset is clearly biased by the last
measurement, but could be eliminated with a $\delta{p} <$~1\% criterion. 
This entire reduction process has been automated using the ORAC-DR
data reduction pipeline \citep{1998adass...7..196E,1999adass...8..171J}.

\section{Scientific capability}

Science targets so far observed include asteroids, planetary nebulae,
supernova remnants, T Tauri star disks, accreting protostars, pre-stellar
cores, Bok globules, dark cloud filaments, high-mass star-forming clouds, the
Galactic Centre including the black hole Sgr A*, the starburst galaxy M82, the
ultraluminous galaxy Arp~220 and a number of variable AGN or `blazars'.
Results have been discussed by
\citet[]{1999ApJ...525..832T,2002ApJ...574..822M,2000ApJ...530L.115D,
2000ApJ...537L.135W,2001ApJ...561..871H,2000ApJ...542..352V,2000AAS...197.0514F,
2000ApJ...534L.173A,2000Natur.404..732G,1999AAS...195.8902M}, among others.
Many of these objects have sub-Jy fluxes per beam, and were not feasible to
observe before SCUBA; they are also very difficult for other ground-based
instruments such as the HERTZ polarimeter-camera used at the CSO, which
operates in the more challenging 350~\mum\ band \citep{1998ApJ...504..588D}.
Examples of some typical observations are described below; the imaging results
shown in the figures have not previously been published.

\subsection{Polarimetric Photometry}

\begin{figure}
\includegraphics[width=84mm]{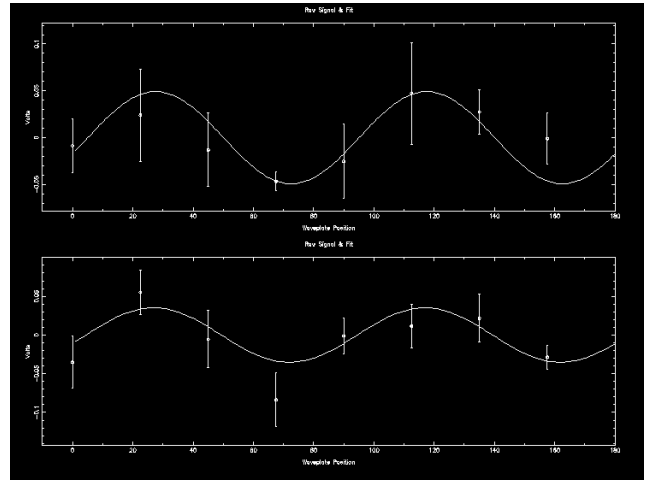}
\caption{Fits to 2000~\mum\ photometric polarimetry of Sgr A*, illustrating
data quality for one waveplate cycle on an 8~Jy, 3\% polarized source 
(instrumental polarization of 1.4\% has not yet been subtracted). The
axes are measured signal modulation versus waveplate orientation, and the 
scales are polarization fraction (tick marks at intervals of 0.01, i.e. 
1\%) versus angle (tick marks at 10$^{\circ}$ intervals). The top plot  
covers $\delta = 0^{\circ}-157.5^{\circ}$ and the bottom plot is for $\delta = 
180^{\circ}-337.5^{\circ}$. Good consistency is demonstrated by the upper and 
lower fits from the first and second halves of the waveplate cycle being in 
phase.  }
\label{fig:phot} 
\end{figure} 

Sgr A*, the massive black hole candidate at the centre of the Galaxy, was
observed polarimetrically by \citet{2000ApJ...534L.173A}. Crucial observations
were made at 1350 and 2000~\mum\ in the photometric mode, to establish the
polarization spectrum in the millimetre regime as well as at submillimetre
wavelengths. These were the first detections of linear polarization of Sgr~A*,
and they showed a dramatic $\sim 90^{\circ}$ shift in position angle around 1
mm wavelength, which implies self-absorption in a very compact emission region
of only a few Schwarzschild radii
\citep[]{2000ApJ...534L.173A,2001ApJ...555L..83B,2002ApJ...573L..23L}. The
1350 and 2000~\mum\ detections were each significant at the 10 sigma level
after about 15 minutes of integration. An example of the fitted results from
one waveplate cycle (2 minutes of data) is shown in Figure~\ref{fig:phot}.

It is now routinely possible to detect polarized emission from compact radio
sources. A $\sim5\sigma$ detection of a 0.5--10 Jy source can be made in
30--60 minutes, depending on the fractional polarization (values of 2--40\%
have been observed). Large levels of polarization indicate that the magnetic
field must be quite highly ordered, and allow its direction to be inferred
without worrying about opacity effects. Comparisons with quasi-simultaneous
results from 7~mm very long base line interferometry show that, in many
cases, the angle of submillimetre polarization matches that very close to the
base of the sub parsec-scale jet, and/or the structural position angle of the
jet (Stevens et al. in prep.).

One of the faintest photometric detections with the UKT14 polarimeter was the
HH24MMS protostar observed by \citet*{1997ApJ...480..255G} at 800~\mum. This
result was confirmed with the SCUBA polarimeter at 850~\mum: HH24MMS was 3.7
$\pm$ 1.3\% polarized at 95 $\pm$ 10$^{\circ}$ in the earlier data and the new
measurement was 3.82 $\pm$ 0.61\% at 109 $\pm$ 5$^{\circ}$. Thus results
published with the old instrument appear to be quite reliable, but now much
fainter targets can be detected.

\subsection{Polarimetric Imaging of Single Fields}

\begin{figure}
\includegraphics[angle=-90,width=84mm,clip=]{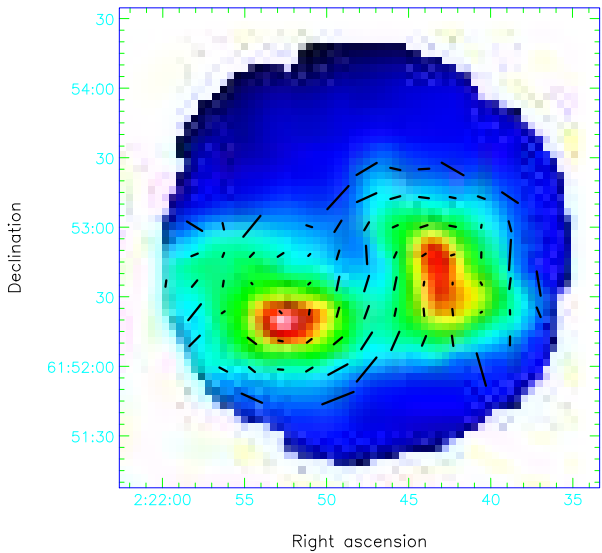}
\caption{850~\mum\ polarization map of the main cores in W3 (see text). The
colour scale goes up to 8 Jy/beam and the vectors range from 0.4\% to 7.6\%;
co-ordinates are B1950. All detections have $p/\delta p > 3$ and $\delta p
< 1$\%. The $E$-plane of polarization is plotted, i.e. parallel
to the long axes of aligned dust grains and perpendicular to the aligning
magnetic field.}
\label{fig:w3} 
\end{figure} 

Jiggle map polarimetry has been used for a wide variety of targets:  
Figure~\ref{fig:w3} illustrates results for the high-mass star-formation
region W3. The two bright cores contain the luminous clusters of young stellar
objects IRS5 and IRS4 (left and right respectively), plus an extended
envelope. The image illustrates the care that must be taken to chop in the
direction of the faintest extended emission (here at a position angle of
$20^{\circ}$ east-of-north and 2.5$'$ throw); there are some minor
discrepancies with the 350\mum\ polarization map of
\citet{2000ApJ...535..913S}, which can plausibly be explained by their larger
(and more suitable) $6.3'$ chop throw. Examination of a JCMT archival
850~\mum\ scan-map shows that the faintest regions with plotted vectors in
Figure~\ref{fig:w3} have in the worst cases up to 25\% relative flux in one of
the off-beams. Although no large-scale polarization data are available, we can
estimate that a 10\% polarization at this off-point would seriously
contaminate a 2.5\% on-source vector. 
  
Projects on regions such as W3 are producing very valuable information on
different magnetic morphologies in clouds with more or less active star
formation and varying core and star masses
\citep[]{2002ApJ...574..822M,2000ApJ...531..868M,2000A&A...356.1031C,
2000ApJ...530L.115D,2000ApJ...544..830F}. Also, the marked decrease in
polarization with increasing flux can be used to test models of field
lines that are tangled or turbulent within the beam, and/or models of
grains that are less well aligned in dense cores
\citep[]{2001ApJ...546..980O,1998ApJ...499L..93A}.

The faintest targets detected with jiggle map polarimetry have fluxes of about
0.2 Jy/beam and a few percent polarization. This covers a wide range of
Galactic interstellar clouds and also the closest star-forming galaxies
\citep{2000Natur.404..732G}. Observing efficiency compared to previous
point-by-point magnetic field mapping has been greatly increased: for example,
the 800\mum\ polarimetry of the W3-IRS4 core reported by
\citet{1999A&A...344..668G} required one hour of observation per spatial
point, whereas the data in Figure~\ref{fig:w3} took a similar time for 78
detected points going down to much lower flux limits.

\subsection{Polarimetric Scan-mapping}

\begin{figure}
\includegraphics[width=84mm,clip=]{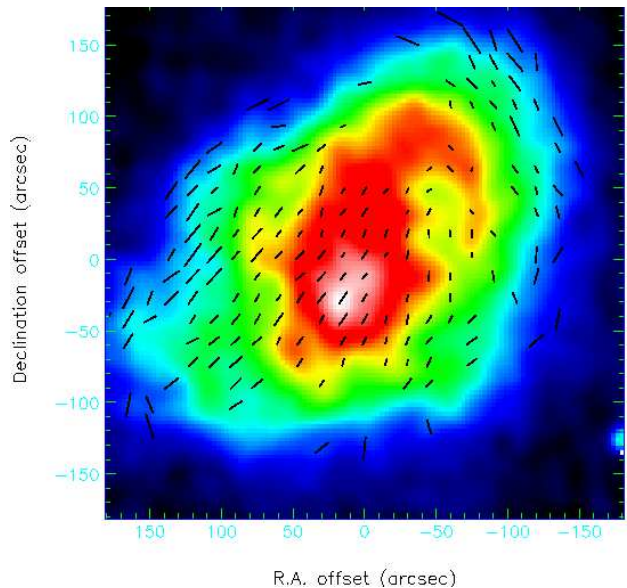}
\caption{850~\mum\ polarization data for the Crab Nebula, using scan map
polarimetry. The smallest and largest polarization vectors are 8\% and 58\%,
and the colour scale goes up to 1.8 Jy/beam at 850~\mum. There are
183 vectors with $p/\delta{p} > 5$, at beam width spacing, after 1 hour of
integration.}
\label{fig:crab}
\end{figure} 

Scan map polarimetry is still under development, but initial tests on the
polarized synchrotron emission of the Crab Nebula gave very good results
(Figure~\ref{fig:crab}). The 850~\mum\ polarization was very similar
between scan map observations and a standard jiggle map (of the central 3
arcmin); percentages and angles differed by only 1\% and 6$^{\circ}$ (the
latter is probably dominated by rotation of a few degrees during each 30
minute map sequence). The angles also agree very well with 9~mm results
published by \citet{1979MNRAS.189..867F}. However, the SCUBA data improve
our understanding of the (sub)millimetre polarization pattern from 90 to
15 arcsecond resolution, greatly clarifying the geometry around features
such as the emission `arm' to the north-west. The average flux and
polarization for the Crab Nebula are 1 Jy/beam and 25\%, and $\geq
5\sigma$ detections were obtained in one hour of observing. A more typical
source with an equivalent polarized flux, such as a 5 Jy/beam, 5\%
polarized dust cloud, should be detectable in the same time.

\section{Current and Future Developments}

Work is ongoing on automating scan-map polarimetry and expanding the use
of 450~\mum\ polarimetry. The latter will give a factor of two improvement
in spatial resolution over 850~\mum\ but not all the IP data have been
obtained.  Fast corrections for atmospheric changes are also being tested,
utilising a water vapour radiometer \citep{2001ApJ...553.1036W} in the
JCMT's receiver cabin. This monitors the profile of the 183 GHz
atmospheric absorption line about every second, and thus gives a very
precise transmission measurement along the same line-of-sight as the
source.  Some tests have also been made of a `continuous spinning' mode,
as an alternative to stepping the waveplate to discrete angles. By
analysing the power spectrum of the signal, it is possible to detect the
intensity of the polarized signal (a modulation every 90$^{\circ}$), and
the position angle can also be measured by retaining information on 
orientation as a function of time. This should be a very fast method of
detecting strong sources, and rapid spinning (e.g. $> 1$~Hz) would
eliminate most problems with the sky changing before the complete
polarized modulation has been detected.

Finally, a replacement camera is currently being built by the UK Astronomy
Technology Centre.  With SCUBA-2 and a new polarimeter, there should be a
factor of 2.5 decrease in NEFD, a factor of 16 increase in imaging speed due
to instantaneous full-sampling (no jiggling) and a 15-fold increase in
field-of-view, combining to give an improvement in polarization imaging power
for large sources of $\sim 1500$.

\section{Summary}

An efficient polarimeter coupled with a very sensitive submillimetre camera has
facilitated a major advance in ground-based imaging polarimetry. The variety of
science targets has also expanded, with almost every category of Galactic
source of dust and synchrotron emission becoming available, as well as a number
of extragalactic objects. Investigations of the dust polarization spectrum can
also be made, especially in conjunction with 60--350~\mum\ polarization data
from the Kuiper Airborne Observatory and the CSO
\citep[]{2000PASP..112.1215H,2000ApJS..128..335D}. This reveals unique
information about the relation of grain alignability and emissivity,
particularly in mixed grain populations
\citep[]{1999A&A...344..668G,1999ApJ...516..834H}. Finally the multi-wavelength
capability of the SCUBA Polarimeter covers both the high-frequency end of
synchrotron spectra and the long-wavelength tail of dust spectra, where both
effects are generally optically thin and easiest to interpret.

\section*{ACKNOWLEDGEMENTS}

The JCMT is operated by the Joint Astronomy Centre on behalf of PPARC of the
UK, the Netherlands OSR, and NRC Canada.  We acknowledge the software and
support provided by the Starlink Project which is run by CCLRC on behalf of
PPARC. We especially wish to thank the many members of JCMT staff involved
in commissioning and maintaining the SCUBA Polarimeter.


\begin{thebibliography}{}

\bibitem[\protect\citeauthoryear{{Aitken}, {Greaves}, {Chrysostomou},
  {Jenness}, {Holland}, {Hough}, {Pierce-Price} \& {Richer}}{{Aitken}
  et~al.}{2000}]{2000ApJ...534L.173A}
{Aitken} D.~K.,  {Greaves} J.,  {Chrysostomou} A.,  {Jenness} T.,  {Holland}
  W.,  {Hough} J.~H.,  {Pierce-Price} D.,    {Richer} J.,  2000, \apj, 534,
  L173

\bibitem[\protect\citeauthoryear{{Arce}, {Goodman}, {Bastien}, {Manset} \&
  {Sumner}}{{Arce} et~al.}{1998}]{1998ApJ...499L..93A}
{Arce} H.~G.,  {Goodman} A.~A.,  {Bastien} P.,  {Manset} N.,    {Sumner} M.,
  1998, \apj, 499, L93

\bibitem[\protect\citeauthoryear{Berry \& Gledhill}{Berry \&
  Gledhill}{2000}]{sun223}
Berry D.~S.,  Gledhill T.~M.,  2000, POLPACK.
Starlink User Note 223, Starlink Project, CLRC

\bibitem[\protect\citeauthoryear{{Bromley}, {Melia} \& {Liu}}{{Bromley}
  et~al.}{2001}]{2001ApJ...555L..83B}
{Bromley} B.~C.,  {Melia} F.,    {Liu} S.,  2001, \apj, 555, L83

\bibitem[\protect\citeauthoryear{{Clemens}, {Kane}, {Leach} \&
  {Barvainis}}{{Clemens} et~al.}{1990}]{1990PASP..102.1064C}
{Clemens} D.~P.,  {Kane} B.~D.,  {Leach} R.~W.,    {Barvainis} R.,  1990,
  \pasp, 102, 1064

\bibitem[\protect\citeauthoryear{{Coppin}, {Greaves}, {Jenness} \&
  {Holland}}{{Coppin} et~al.}{2000}]{2000A&A...356.1031C}
{Coppin} K.~E.~K.,  {Greaves} J.~S.,  {Jenness} T.,    {Holland} W.~S.,  2000,
  \aap, 356, 1031

\bibitem[\protect\citeauthoryear{{Davis}, {Chrysostomou}, {Matthews}, {Jenness}
  \& {Ray}}{{Davis} et~al.}{2000}]{2000ApJ...530L.115D}
{Davis} C.~J.,  {Chrysostomou} A.,  {Matthews} H.~E.,  {Jenness} T.,    {Ray}
  T.~P.,  2000, \apj, 530, L115

\bibitem[\protect\citeauthoryear{{Dotson}, {Davidson}, {Dowell}, {Schleuning}
  \& {Hildebrand}}{{Dotson} et~al.}{2000}]{2000ApJS..128..335D}
{Dotson} J.~L.,  {Davidson} J.,  {Dowell} C.~D.,  {Schleuning} D.~A.,
  {Hildebrand} R.~H.,  2000, \apjs, 128, 335

\bibitem[\protect\citeauthoryear{{Dowell}, {Hildebrand}, {Schleuning},
  {Vaillancourt}, {Dotson}, {Novak}, {Renbarger} \& {Houde}}{{Dowell}
  et~al.}{1998}]{1998ApJ...504..588D}
{Dowell} C.~D.,  {Hildebrand} R.~H.,  {Schleuning} D.~A.,  {Vaillancourt}
  J.~E.,  {Dotson} J.~L.,  {Novak} G.,  {Renbarger} T.,    {Houde} M.,  1998,
  \apj, 504, 588+

\bibitem[\protect\citeauthoryear{{Economou}, {Bridger}, {Wright}, {Rees} \&
  {Jenness}}{{Economou} et~al.}{1998}]{1998adass...7..196E}
{Economou} F.,  {Bridger} A.,  {Wright} G.~S.,  {Rees} N.~P.,    {Jenness} T.,
  1998, in ASP Conf. Ser. 145: Astronomical Data Analysis Software and Systems
  VII. pp~196+

\bibitem[\protect\citeauthoryear{{Feldman}, {Redman}, {Carey} \&
  {Egan}}{{Feldman} et~al.}{2000}]{2000AAS...197.0514F}
{Feldman} P.~A.,  {Redman} R.~O.,  {Carey} S.~J.,    {Egan} M.~P.,  2000, in
  American Astronomical Society Meeting. pp 0514+

\bibitem[\protect\citeauthoryear{{Fiege} \& {Pudritz}}{{Fiege} \&
  {Pudritz}}{2000}]{2000ApJ...544..830F}
{Fiege} J.~D.,  {Pudritz} R.~E.,  2000, \apj, 544, 830

\bibitem[\protect\citeauthoryear{{Flett} \& {Henderson}}{{Flett} \&
  {Henderson}}{1979}]{1979MNRAS.189..867F}
{Flett} A.~M.,  {Henderson} C.,  1979, \mnras, 189, 867

\bibitem[\protect\citeauthoryear{{Flett} \& {Murray}}{{Flett} \&
  {Murray}}{1991}]{1991MNRAS.249P...4F}
{Flett} A.~M.,  {Murray} A.~G.,  1991, \mnras, 249, 4P

\bibitem[\protect\citeauthoryear{{Glenn}, {Walker} \& {Young}}{{Glenn}
  et~al.}{1999}]{1999ApJ...511..812G}
{Glenn} J.,  {Walker} C.~K.,    {Young} E.~T.,  1999, \apj, 511, 812

\bibitem[\protect\citeauthoryear{Greaves}{Greaves}{2002}]{2002A&A..PNE}
Greaves J.~S.,  2002, \aap, in press

\bibitem[\protect\citeauthoryear{{Greaves}, {Holland}, {Jenness} \&
  {Hawarden}}{{Greaves} et~al.}{2000}]{2000Natur.404..732G}
{Greaves} J.~S.,  {Holland} W.~S.,  {Jenness} T.,    {Hawarden} T.~G.,  2000,
  \nat, 404, 732

\bibitem[\protect\citeauthoryear{{Greaves}, {Holland}, {Minchin}, {Murray} \&
  {Stevens}}{{Greaves} et~al.}{1999}]{1999A&A...344..668G}
{Greaves} J.~S.,  {Holland} W.~S.,  {Minchin} N.~R.,  {Murray} A.~G.,
  {Stevens} J.~A.,  1999, \aap, 344, 668

\bibitem[\protect\citeauthoryear{{Greaves}, {Holland} \&
  {Ward-Thompson}}{{Greaves} et~al.}{1997}]{1997ApJ...480..255G}
{Greaves} J.~S.,  {Holland} W.~S.,    {Ward-Thompson} D.,  1997, \apj, 480,
  255+

\bibitem[\protect\citeauthoryear{{Henning}, {Wolf}, {Launhardt} \&
  {Waters}}{{Henning} et~al.}{2001}]{2001ApJ...561..871H}
{Henning} T.,  {Wolf} S.,  {Launhardt} R.,    {Waters} R.,  2001, \apj, 561,
  871

\bibitem[\protect\citeauthoryear{{Hildebrand}, {Davidson}, {Dotson}, {Dowell},
  {Novak} \& {Vaillancourt}}{{Hildebrand} et~al.}{2000}]{2000PASP..112.1215H}
{Hildebrand} R.~H.,  {Davidson} J.~A.,  {Dotson} J.~L.,  {Dowell} C.~D.,
  {Novak} G.,    {Vaillancourt} J.~E.,  2000, \pasp, 112, 1215

\bibitem[\protect\citeauthoryear{{Hildebrand}, {Dotson}, {Dowell}, {Schleuning}
  \& {Vaillancourt}}{{Hildebrand} et~al.}{1999}]{1999ApJ...516..834H}
{Hildebrand} R.~H.,  {Dotson} J.~L.,  {Dowell} C.~D.,  {Schleuning} D.~A.,
  {Vaillancourt} J.~E.,  1999, \apj, 516, 834

\bibitem[\protect\citeauthoryear{{Holland}, {Cunningham}, {Gear}, {Jenness},
  {Laidlaw}, {Lightfoot} \& {Robson}}{{Holland}
  et~al.}{1998}]{1998SPIE.3357..305H}
{Holland} W.~S.,  {Cunningham} C.~R.,  {Gear} W.~K.,  {Jenness} T.,  {Laidlaw}
  K.,  {Lightfoot} J.~F.,    {Robson} E.~I.,  1998, in Proc. SPIE Vol. 3357, p.
  305-318, Advanced Technology MMW, Radio, and Terahertz Telescopes, Thomas G.
  Phillips; Ed.. pp 305--318

\bibitem[\protect\citeauthoryear{{Holland}, {Robson}, {Gear}, {Cunningham},
  {Lightfoot}, {Jenness}, {Ivison}, {Stevens}, {Ade}, {Griffin}, {Duncan},
  {Murphy} \& {Naylor}}{{Holland} et~al.}{1999}]{1999MNRAS.303..659H}
{Holland} W.~S.,  {Robson} E.~I.,  {Gear} W.~K.,  {Cunningham} C.~R.,
  {Lightfoot} J.~F.,  {Jenness} T.,  {Ivison} R.~J.,  {Stevens} J.~A.,  {Ade}
  P. A.~R.,  {Griffin} M.~J.,  {Duncan} W.~D.,  {Murphy} J.~A.,    {Naylor}
  D.~A.,  1999, \mnras, 303, 659

\bibitem[\protect\citeauthoryear{{Jenness} \& {Economou}}{{Jenness} \&
  {Economou}}{1999}]{1999adass...8..171J}
{Jenness} T.,  {Economou} F.,  1999, in Mehringer D.~M.,  Plante R.~L.,
  Roberts D.~A.,  eds,  ASP Conf. Ser. Vol. 172, Astronomical Data Analysis
  Software and Systems VIII. Astron. Soc. Pac., San Francisco, p.~171

\bibitem[\protect\citeauthoryear{{Jenness}, {Holland}, {Chapin}, {Lightfoot} \&
  {Duncan}}{{Jenness} et~al.}{2000}]{2000adass...9..559J}
{Jenness} T.,  {Holland} W.~S.,  {Chapin} E.,  {Lightfoot} J.~F.,    {Duncan}
  W.~D.,  2000, in ASP Conf. Ser. 216: Astronomical Data Analysis Software and
  Systems IX. pp~559+

\bibitem[\protect\citeauthoryear{Jenness \& Lightfoot}{Jenness \&
  Lightfoot}{1998}]{1998adass...7..216J}
Jenness T.,  Lightfoot J.~F.,  1998, in Albrecht R.,  Hook R.~N.,   Bushouse
  H.~A.,  eds,  ASP Conf. Ser. Vol. 145, Astronomical Data Analysis Software
  and Systems VII. Astron. Soc. Pac., San Francisco, p.~216

\bibitem[\protect\citeauthoryear{{Jenness}, {Lightfoot} \& {Holland}}{{Jenness}
  et~al.}{1998}]{1998SPIE.3357..548J}
{Jenness} T.,  {Lightfoot} J.~F.,    {Holland} W.~S.,  1998, \procspie, 3357,
  548

\bibitem[\protect\citeauthoryear{{Jenness}, {Lightfoot}, {Holland}, {Greaves}
  \& {Economou}}{{Jenness} et~al.}{2000}]{2000irsm.conf..205J}
{Jenness} T.,  {Lightfoot} J.~F.,  {Holland} W.~S.,  {Greaves} J.~S.,
  {Economou} F.,  2000, in ASP Conf. Ser. 217: Imaging at Radio through
  Submillimeter Wavelengths. pp~205+

\bibitem[\protect\citeauthoryear{Jenness, Stevens, Archibald, Economou, Jessop
  \& Robson}{Jenness et~al.}{2002}]{2002MNRAS..TJCAL}
Jenness T.,  Stevens J.,  Archibald E.~N.,  Economou F.,  Jessop N.~E.,
  Robson E.~I.,  2002, \mnras, 336, 14

\bibitem[\protect\citeauthoryear{{Kane}, {Clemens}, {Barvainis} \&
  {Leach}}{{Kane} et~al.}{1993}]{1993ApJ...411..708K}
{Kane} B.~D.,  {Clemens} D.~P.,  {Barvainis} R.,    {Leach} R.~W.,  1993, \apj,
  411, 708

\bibitem[\protect\citeauthoryear{{Liu} \& {Melia}}{{Liu} \&
  {Melia}}{2002}]{2002ApJ...573L..23L}
{Liu} S.,  {Melia} F.,  2002, \apj, 573, L23

\bibitem[\protect\citeauthoryear{{Marscher}, {Marchenko}, {Stevens}, {Gear},
  {Lister}, {Cawthorne}, {Stirling}, {G{\' o}mez}, {Gabuzda} \&
  {Robson}}{{Marscher} et~al.}{1999}]{1999AAS...195.8902M}
{Marscher} A.~P.,  {Marchenko} S.~G.,  {Stevens} J.~A.,  {Gear} W.~K.,
  {Lister} M.~L.,  {Cawthorne} T.~V.,  {Stirling} A.,  {G{\' o}mez} J.~L.,
  {Gabuzda} D.~C.,    {Robson} E.~I.,  1999, in American Astronomical Society
  Meeting. pp 8902+

\bibitem[\protect\citeauthoryear{{Matthews} \& {Wilson}}{{Matthews} \&
  {Wilson}}{2000}]{2000ApJ...531..868M}
{Matthews} B.~C.,  {Wilson} C.~D.,  2000, \apj, 531, 868

\bibitem[\protect\citeauthoryear{{Matthews} \& {Wilson}}{{Matthews} \&
  {Wilson}}{2002}]{2002ApJ...574..822M}
{Matthews} B.~C.,  {Wilson} C.~D.,  2002, \apj, 574, 822

\bibitem[\protect\citeauthoryear{{Matthews}, {Wilson} \& {Fiege}}{{Matthews}
  et~al.}{2001}]{2001ApJ...562..400M}
{Matthews} B.~C.,  {Wilson} C.~D.,    {Fiege} J.~D.,  2001, \apj, 562, 400

\bibitem[\protect\citeauthoryear{{Murray}, {Nartallo}, {Haynes}, {Gannaway} \&
  {Ade}}{{Murray} et~al.}{1997}]{1997fisu.conf..405M}
{Murray} A.~G.,  {Nartallo} R.,  {Haynes} C.~V.,  {Gannaway} F.,    {Ade}
  P.~A.~R.,  1997, in The Far Infrared and Submillimetre Universe.. pp~405+

\bibitem[\protect\citeauthoryear{{Naghizadeh-Khouei} \&
  {Clarke}}{{Naghizadeh-Khouei} \& {Clarke}}{1993}]{1993A&A...274..968N}
{Naghizadeh-Khouei} J.,  {Clarke} D.,  1993, \aap, 274, 968+

\bibitem[\protect\citeauthoryear{{Nartallo}}{{Nartallo}}{1995}]{1995PhDT......%
..54N}
{Nartallo} R.,  1995, Ph.D.~Thesis, pp~54+

\bibitem[\protect\citeauthoryear{{Nartallo}, {Gear}, {Murray}, {Robson} \&
  {Hough}}{{Nartallo} et~al.}{1998}]{1998MNRAS.297..667N}
{Nartallo} R.,  {Gear} W.~K.,  {Murray} A.~G.,  {Robson} E.~I.,    {Hough}
  J.~H.,  1998, \mnras, 297, 667

\bibitem[\protect\citeauthoryear{{Novak}, {Gonatas}, {Hildebrand}, {Platt} \&
  {Dragovan}}{{Novak} et~al.}{1989}]{1989ApJ...345..802N}
{Novak} G.,  {Gonatas} D.~P.,  {Hildebrand} R.~H.,  {Platt} S.~R.,
  {Dragovan} M.,  1989, \apj, 345, 802

\bibitem[\protect\citeauthoryear{{Ostriker}, {Stone} \& {Gammie}}{{Ostriker}
  et~al.}{2001}]{2001ApJ...546..980O}
{Ostriker} E.~C.,  {Stone} J.~M.,    {Gammie} C.~F.,  2001, \apj, 546, 980

\bibitem[\protect\citeauthoryear{{Schleuning}}{{Schleuning}}{1998}]{1998ApJ...%
493..811S}
{Schleuning} D.~A.,  1998, \apj, 493, 811+

\bibitem[\protect\citeauthoryear{{Schleuning}, {Vaillancourt}, {Hildebrand},
  {Dowell}, {Novak}, {Dotson} \& {Davidson}}{{Schleuning}
  et~al.}{2000}]{2000ApJ...535..913S}
{Schleuning} D.~A.,  {Vaillancourt} J.~E.,  {Hildebrand} R.~H.,  {Dowell}
  C.~D.,  {Novak} G.,  {Dotson} J.~L.,    {Davidson} J.~A.,  2000, \apj, 535,
  913

\bibitem[\protect\citeauthoryear{{Tamura}, {Hough}, {Greaves}, {Morino},
  {Chrysostomou}, {Holland} \& {Momose}}{{Tamura}
  et~al.}{1999}]{1999ApJ...525..832T}
{Tamura} M.,  {Hough} J.~H.,  {Greaves} J.~S.,  {Morino} J.,  {Chrysostomou}
  A.,  {Holland} W.~S.,    {Momose} M.,  1999, \apj, 525, 832

\bibitem[\protect\citeauthoryear{{Title} \& {Rosenberg}}{{Title} \&
  {Rosenberg}}{1981}]{1981SPIE.307..120T}
{Title} A.~M.,  {Rosenberg} W.~J.,  1981, in Proc. SPIE Vol. 307, p. 120+,
  Polarizers and Applications, G.~B. Trapani; Ed.. pp~120+

\bibitem[\protect\citeauthoryear{{Vall{\' e}e}, {Bastien} \&
  {Greaves}}{{Vall{\' e}e} et~al.}{2000}]{2000ApJ...542..352V}
{Vall{\' e}e} J.~P.,  {Bastien} P.,    {Greaves} J.~S.,  2000, \apj, 542, 352

\bibitem[\protect\citeauthoryear{{Ward-Thompson}, {Kirk}, {Crutcher},
  {Greaves}, {Holland} \& {Andr{\' e}}}{{Ward-Thompson}
  et~al.}{2000}]{2000ApJ...537L.135W}
{Ward-Thompson} D.,  {Kirk} J.~M.,  {Crutcher} R.~M.,  {Greaves} J.~S.,
  {Holland} W.~S.,    {Andr{\' e}} P.,  2000, \apj, 537, L135

\bibitem[\protect\citeauthoryear{{Wiedner}, {Hills}, {Carlstrom} \&
  {Lay}}{{Wiedner} et~al.}{2001}]{2001ApJ...553.1036W}
{Wiedner} M.~C.,  {Hills} R.~E.,  {Carlstrom} J.~E.,    {Lay} O.~P.,  2001,
  \apj, 553, 1036

\end{thebibliography}

\bsp

\end{document}